\documentclass[12pt]{article}
\usepackage{epsf}
\usepackage{amsmath, amssymb}

\begin{document}

\renewcommand{\thefootnote}{\fnsymbol{footnote}}
\setcounter{footnote}{0}
\begin{titlepage}

\def\thefootnote{\fnsymbol{footnote}}

\begin{center}

\hfill TU-858\\
\hfill ICRR-Report-555-2009-17\\
\hfill IPMU 09-0144  \\
\hfill December, 2009\\

\vskip .75in

{\Large \bf 
Right-handed sneutrino dark matter\\
and big-bang nucleosynthesis
}

\vskip .75in

{\large 
Koji Ishiwata$^{(a)}$, Masahiro Kawasaki$^{(b,c)}$, Kazunori
Kohri$^{(a)}$, \\ and Takeo Moroi$^{(a,c)}$ }

\vskip 0.25in

{\em

$^a$Department of Physics, Tohoku University,
Sendai 980-8578, Japan

$^b$Institute for Cosmic Ray Research, University of Tokyo, Kashiwa
277-8582, Japan

$^c$Institute for the Physics and Mathematics of the Universe,
 University of Tokyo,\\
Kashiwa 277-8568, Japan
}

\end{center}
\vskip .5in

\begin{abstract}

  We study the light-element abundances in supersymmetric model where
  the right-handed sneutrino is the lightest superparticle (LSP),
  assuming that the neutrino masses are purely Dirac-type.  In such a
  scenario, the lightest superparticle in the minimal supersymmetric
  standard model sector (which we call MSSM-LSP) becomes long-lived,
  and thermal relic MSSM-LSP may decay after the big-bang
  nucleosynthesis starts.  We calculate the light-element abundances
  including non-standard nuclear reactions induced by the MSSM-LSP
  decay, and derive constraints on the scenario of right-handed
  sneutrino LSP.

\end{abstract}

\end{titlepage}

\renewcommand{\thepage}{\arabic{page}}
\setcounter{page}{1}
\renewcommand{\thefootnote}{\#\arabic{footnote}}
\setcounter{footnote}{0}

With the precise astrophysical observations, it is now widely believed
that about $23 \%$ of the energy density of the present universe is
due to dark matter (DM) \cite{Hinshaw:2008kr}.  The existence of dark
matter, however, raises a serious question to particle physics because
there is no viable candidate for dark matter in the particle content
of the standard model.  To solve this problem, many dark-matter models
have been proposed so far.

In constructing dark-matter model, it is important to understand how
dark matter was produced in the early universe.  In many cases, the
thermal freeze-out mechanism is adopted to produce dark matter
particle in the early universe; then, dark matter particle, which is
in thermal bath when the cosmic temperature is higher than its mass,
freezes out from the thermal bath when the cosmic temperature becomes
low.

However, the freeze-out scenario is not the only possibility to
produce dark matter particle in the early universe.  Even if the dark
matter particle is very weakly interacting so that it is never
thermalized, it can be produced by the decay and scattering of
particles in thermal bath.  In particular, if the interaction of dark
matter is dominated by renormalizable ones, dark-matter production is
most effective when the temperature is comparable to the mass of
parent particle which produces dark matter via the decay or
scattering.  Thus, if the reheating temperature after inflation is
higher than the mass of parent particle, the relic density of the dark
matter becomes insensitive to the cosmic evolution in the early stage.

Such a scenario was originally proposed in \cite{SnurLSP}, where the
right-handed sneutrino $\tilde{\nu}_R$ in supersymmetric model is
shown to be a viable candidate for dark matter.  In \cite{SnurLSP}, it
was also shown that, if $\tilde{\nu}_R$-DM is dominantly produced from
the decay and scattering of superparticles in thermal bath, the
primordial abundance of $\tilde{\nu}_R$ is determined when the cosmic
temperature is comparable to the masses of superparticles.  Then,
recently, more general discussion of such a scenario has been given in
\cite{Hall:2009bx}, where a variety of candidates for such very
weakly-interacting dark-matter particles have been also considered.

If a very weakly interacting particle is dark matter, it is often the
case that a long-lived particle (with lifetime longer than $\sim 1\
{\rm sec}$) may show up, which may spoil the success of the standard
big-bang nucleosynthesis \cite{Moroi:1993mb, SuperWIMP, SnurLSP,
  Ishiwata:2007bt, Hall:2009bx}.  This is indeed the case where the
right-handed sneutrino is the lightest superparticle (LSP) and is dark
matter.  If a right-handed sneutrino is the LSP, the lightest
superparticle in the minimal supersymmetric standard model (MSSM)
sector (which we call MSSM-LSP) decays into $\tilde{\nu}_R$ (and
$R$-even particles) via very small neutrino Yukawa interaction.  Then,
decay of the MSSM-LSP after the big-bang nucleosynthesis (BBN) epoch
may affect the light-element abundances.  Thus, it is important to
check the BBN constraints on the scenario.

In this letter, we consider the case where a right-handed sneutrino is
the LSP, assuming Dirac-type neutrino masses \cite{SnurLSP}.\footnote
{For related topics, see also \cite{VariousSnuDMPapaers}.}
We study the light-element abundances in such a case in detail, and
derive BBN constraints on the mass and lifetime of the MSSM-LSP.  We
also comment on the implication of the sneutrino LSP scenario on the
${\rm ^7Li}$ overproduction problem.

First, we discuss the model framework that we consider in this letter.
The superpotential is written as
\begin{eqnarray}
 W = W_{\rm MSSM} 
   +y_{\nu} \hat{L} \hat{H}_u \hat{\nu}^c_R,
\end{eqnarray}
where $W_{\rm MSSM}$ is the superpotential of the MSSM, $\hat{L} =
(\hat{\nu}_L, \hat{e}_L)$ and $\hat{H}_u=(\hat{H}^+_u, \hat{H}^0_u)$
are left-handed lepton doublet and up-type Higgs doublet,
respectively.  (In this letter, ``hat'' is used for superfields, while
``tilde'' is for superpartners.)  Generation indices are omitted for
simplicity.  In this model, neutrinos acquire their masses only
through Yukawa interactions as $m_{\nu} = y_{\nu} \langle H^0_{u}
\rangle = y_{\nu} v \sin{\beta}$, where $v \simeq$ 174 GeV is the
vacuum expectation value (VEV) of the standard model Higgs field and
$\tan{\beta} = \langle H^0_u \rangle/\langle H^0_d \rangle$. Thus, the
neutrino Yukawa coupling is determined by the neutrino mass as
\begin{eqnarray}
 y_{\nu} \sin{\beta}
 =
 3.0 \times 10^{-13}
 \times 
 \left(
  \frac{ m^2_{\nu} }{ 2.8 \times 10^{-3} ~{\rm eV}^2 }
 \right)^{1/2}.
\end{eqnarray}
Mass squared differences among neutrinos have already been determined
accurately by neutrino oscillation experiments.  In particular, the
K2K experiment suggests $[ \Delta m_{\nu}^2 ]_{\rm atom}\simeq (1.9 -
3.5) \times 10^{-3}~{\rm eV^2}$ \cite{Ahn:2006zza}.  In the following
discussion, we assume that the spectrum of neutrino masses is
hierarchical, hence the largest neutrino Yukawa coupling is of the
order of $10^{-13}$ unless otherwise mentioned; we use $y_{\nu}=3.0
\times 10^{-13}$ for our numerical study.  (We neglect effects of
smaller Yukawa coupling constants.)  For our study, it is also
necessary to introduce soft supersymmetry (SUSY) breaking terms.  Soft
SUSY breaking terms relevant to our analysis are
\begin{eqnarray}
  {\cal L}_{\rm soft} 
  = 
  - \frac{1}{2}
  (
  m_{\tilde{B}}\tilde{B}\tilde{B}+m_{\tilde{W}}\tilde{W}\tilde{W}
  + {\rm h.c.} )
  - M^2_{\tilde{L}} \tilde{L}^{\dagger} \tilde{L} 
 - M^2_{\tilde{\nu}_R} \tilde{\nu}^{*}_R   \tilde{\nu}_R
 +( A_{\nu} \tilde{L} H_u \tilde{\nu}^c_R + {\rm h.c.} ),
\end{eqnarray}
where $\tilde{B}$ and $\tilde{W}$ are Bino and Wino, respectively.  We
parameterize $A_{\nu}$ by using the dimensionless constant $a_\nu$ as
\begin{eqnarray}
  A_{\nu} = a_\nu y_{\nu} M_{\tilde{L}}.
  \label{A_nu}
\end{eqnarray}
Notice that $a_\nu$ is a free parameter and, in gravity-mediated SUSY
breaking scenario, for example, $a_\nu$ is expected to be $O(1)$.  The
$A_\nu$-term induces the left-right mixing in the sneutrino mass
matrix, through which the MSSM-LSP decays in the present case.  In the
calculation of mass eigenvalues, however, the mixing is negligible
because of the smallness of neutrino Yukawa coupling constants, and we
obtain
\begin{eqnarray}
  m^2_{\tilde{\nu}_L}
  \simeq M^2_{\tilde{L}} + \frac{1}{2} m^2_Z \cos 2 \beta,
  \quad
  m^2_{\tilde{\nu}_R}
  \simeq M^2_{\tilde{\nu}_R},
  \label{snumass}
\end{eqnarray}
where $m_Z$ is the $Z$ boson mass.  Here and hereafter, we assume that
all the right-handed sneutrinos are degenerate in mass for simplicity.
In the numerical study, we take the following model parameters:
$m_{\tilde{\nu}_R}=100\ {\rm GeV}$, $\tan \beta=30$, and $m_h=115~{\rm
  GeV}$ (with $m_h$ being the lightest Higgs boson mass).  In
addition, the Wino mass is related to the Bino mass using the GUT
relation.

In the early universe, right-handed sneutrino is never thermalized
because of the weakness of neutrino Yukawa interaction.  Although it
is decoupled from thermal bath, right-handed sneutrino can be produced
in various processes; (i) decay or scattering of MSSM particles in
thermal bath, (ii) decay of MSSM-LSP after freeze-out, and (iii)
production in very early universe via the decay of exotic particles
(like gravitino or inflaton).  Thereafter, we donate the contribution
of each process as, $\Omega_{\tilde{\nu}_R}^{\mbox{\tiny (Thermal)}}$,
$\Omega_{\tilde{\nu}_R}^{\mbox{\tiny (F.O.)}}$, and
$\Omega_{\tilde{\nu}_R}^{\mbox{\tiny (non-MSSM)}}$ in order.
Primarily, right-handed sneutrino is produced through neutrino Yukawa
interaction (and the left-right mixing of sneutrino) dominantly in the
following decay processes: $\tilde{H}^0 \rightarrow \tilde{\nu}_R
\bar{\nu}$, $\tilde{H}^+ \rightarrow \tilde{\nu}_R l^+$,
$\tilde{\nu}_L \rightarrow \tilde{\nu}_R h$, $\tilde{\nu}_L\rightarrow
\tilde{\nu}_R Z$, $\tilde{l}_L\rightarrow \tilde{\nu}_R W^-$,
$\tilde{B}\rightarrow \tilde{\nu}_R \bar{\nu}$,
$\tilde{W}^0\rightarrow \tilde{\nu}_R \bar{\nu}$, and $\tilde{W}^+
\rightarrow \tilde{\nu}_R l^+$.  In the previous work, it was shown
that right-handed sneutrino can be adequately produced to become dark
matter when the masses of left- and right-handed sneutrino are
degenerate at $10 - 20$\ \% with $a_{\nu}\lesssim 3$, or in a case of
larger $a_{\nu}$ without degeneracy \cite{SnurLSP}.  It is also
mentioned that enhancement of right-handed sneutrino production is
possible with larger neutrino Yukawa coupling if we consider the case
where neutrino masses are degenerate.  Giving an eye on the thermal
bath again, the MSSM-LSP decouples from thermal bath and its number
freezes out in the same manner with usual MSSM, while the number of
the other MSSM particles is suppressed by Boltzmann factor in this
epoch.  However, relic MSSM-LSP, which is assumed to be the
next-to-the-lightest superparticle (NLSP) in this letter, decays to
right-handed sneutrino through neutrino Yukawa coupling in the late
time.  In this process, the contribution to the abundance is given as
\begin{eqnarray}
  \Omega_{\tilde{\nu}_R}^{\rm (F.O.)} = 
  \frac{m_{\tilde{\nu}_R}}{m_{\rm NLSP}} 
  \Omega_{\rm NLSP}^{\rm (F.O.)},
  \label{Omega(FO)}
\end{eqnarray}
where $m_{\rm NLSP}$ is the mass of the NLSP and $\Omega_{\rm
  NLSP}^{\rm (F.O.)}$ is the would-be density parameter of the relic
NLSP (for the case where it does not decay into $\tilde{\nu}_R$).
Lastly, we mention that there might be a possibility that right-handed
sneutrino is produced directly from an exotic particle in the very
early universe.  The abundance of the expected right-handed sneutrino
is model-dependent, and we do not discuss further detail of specific
model.  In this letter, we consider the scenario that right-handed
sneutrino produced in these processes becomes dark matter.  We do not
specify which is dominant process to produce right-handed sneutrino.

If $\tilde{\nu}_R$ is the LSP, it is always the case that the MSSM-LSP
becomes long-lived.  Because some amount of relic NLSP always exists
in the early universe, they may cause serious problem in BBN; if the
relic MSSM-LSP decays during or after the BBN epoch, energetic charged
and/or colored particles are emitted; they cause the photo- and
hadro-dissociation processes of light elements, which may spoil the
success of the standard BBN scenario.  In the following, we consider
three typical candidates for the MSSM-LSP; Bino $\tilde{B}$,
left-handed sneutrino $\tilde{\nu}_L$, and lighter stau
$\tilde{\tau}$, and study how the $\tilde{\nu}_R$-DM scenario is
constrained by the BBN.

In the Bino-NLSP case, the Bino dominantly decays as
$\tilde{B}\rightarrow \tilde{\nu}_R \bar{\nu}$ (and its CP-conjugated
process) and its decay rate is given by\footnote
{In this letter, we consider the case where the Gaugino-Higgsino
  mixing is so small that its effect is negligible.}
\begin{eqnarray}
    \Gamma_{\tilde{B}\rightarrow \tilde{\nu}_R \bar{\nu}} 
    =
    \frac{\beta_{\rm f}^2 g_1^2}{64\pi}
    \left[\frac{A_\nu v}
        {m_{\tilde{\nu}_L}^2-m_{\tilde{\nu}_R}^2}
    \right]^2
    m_{\tilde{B}},
    \label{Gamma_bino}
\end{eqnarray}
where $g_1$ is the $U(1)_Y$ gauge coupling constant and, for the
process $x\rightarrow \tilde{\nu}_Ry$, $\beta_{\rm f}$ is given by
\begin{eqnarray}
    \beta_{\rm f}^2 = \frac{1}{m_x^4}
    [ m_x^4 - 2 (m_{\tilde{\nu}_R}^2 + m_y^2) m_x^2
    + (m_{\tilde{\nu}_R}^2 - m_y^2)^2 ],
\end{eqnarray}
with $m_x$ and $m_y$ being the masses of the particles $x$ and $y$,
respectively.  When $\tilde{\nu}_L$ or $\tilde{\tau}$ is the NLSP, the
NLSP decays by emitting weak- or Higgs-boson if kinematically allowed.
The decay rates for those processes are given by
\begin{eqnarray}
    \Gamma_{\tilde{\nu}_L\rightarrow \tilde{\nu}_R Z} 
    &=&
    \frac{\beta_{\rm f}^3}{32\pi} 
    \left[\frac{m_{\tilde{\nu}_L}^2}
        {m_{\tilde{\nu}_L}^2-m_{\tilde{\nu}_R}^2}
    \right]^2 
    \frac{A_\nu^2}{m_{\tilde{\nu}_L}},
    \label{Gamma_snul2Z}
    \\ 
    \Gamma_{\tilde{\nu}_L\rightarrow \tilde{\nu}_R h} 
    &=&
    \frac{\beta_{\rm f}}{32\pi} \frac{A_\nu^2}{m_{\tilde{\nu}_L}},
    \label{Gamma_snul2h}
    \\ 
    \Gamma_{\tilde{\tau}\rightarrow \tilde{\nu}_R W^-} 
    &=&
    \frac{\beta_{\rm f}^3 \sin^2\theta_{\tilde{\tau}}}{16\pi} 
    \left[\frac{m_{\tilde{\tau}}^2}
        {m_{\tilde{\nu}_L}^2-m_{\tilde{\nu}_R}^2}
    \right]^2 
    \frac{A_\nu^2}{m_{\tilde{\tau}}},
    \label{Gamma_stau2h}
\end{eqnarray}
where $m_{\tilde{\tau}}$ is the stau mass and $\theta_{\tilde{\tau}}$
is the left-right mixing angle of stau.  (The lighter stau is given by
$\tilde{\tau}=\tilde{\tau}_R\cos\theta_{\tilde{\tau}}
+\tilde{\tau}_L\sin\theta_{\tilde{\tau}}$.)  If the two-body processes
are kinematically blocked, the slepton-NLSP decays into three-body
final state as $\tilde{\nu}_L\rightarrow \tilde{\nu}_R f\bar{f}$ and
$\tilde{\tau}\rightarrow \tilde{\nu}_R f\bar{f}'$ (with $f$ and $f'$
being standard-model fermions).  

Now, we are at the position to discuss the BBN constraints on
$\tilde{\nu}_R$-DM scenario.  We start with the case where Bino is
NLSP.  As we have mentioned, the Bino-NLSP dominantly decays as
$\tilde{B}\rightarrow \tilde{\nu}_R\bar{\nu}$.  Since $\tilde{\nu}_R$
and $\nu$ are (very) weakly interacting particles, the BBN constraints
are not so severe if this is the only possible decay mode.  However,
$\tilde{B}$ may also decay as $\tilde{B}\rightarrow
\tilde{\nu}_R\bar{\nu} Z^{(*)}$ and $\tilde{\nu}_R l W^{(*)}$, where
$Z^{(*)}$ and $W^{(*)}$ are on-shell or off-shell $Z$ and $W$ bosons
(where the ``star'' is for off-shell particle), respectively, while
$l$ is charged lepton.  Then, through the decay of $Z^{(*)}$ and
$W^{(*)}$, quarks and charged leptons are produced.  Even though the
branching ratio for such processes are phase-space suppressed, they
produce sizable amount of hadrons which may significantly affect the
light-element abundances.  Thus, in our analysis, effects of those
decay modes are taken into account in deriving the BBN constraints.
The light-element abundances also depend on the primordial abundance
of the NLSP, and we adopt the abundance of Bino in the focus-point (or
co-annihilation) region \cite{Feng:2004mt}\footnote
{If the Bino is the NLSP, its primordial abundance strongly depends on
  the MSSM parameters.  In the so-called bulk region, the abundance is
  larger, and is approximately given by
  \begin{eqnarray*}
    Y_{\tilde{B}}^{\rm (bulk)} =
    4 \times 10^{-12} \times 
    \left( \frac{m_{\tilde{B}}}{100\ {\rm GeV}} \right).
  \end{eqnarray*}
  We have checked that the BBN constraints in such a case are almost
  the same as the focus-point case.  If we adopt the abundance in the
  bulk region, however, $\Omega_{\tilde{\nu}_R}^{\rm (F.O.)}$ becomes
  larger than the present dark matter density if
  $m_{\tilde{\nu}_R}=100\ {\rm GeV}$.  Thus we will not consider such
  a case in the following discussion.}
\begin{eqnarray}
  Y^{\rm (focus)}_{\tilde{B}} = 
  9 \times 10^{-13} \times 
  \left( \frac{m_{\tilde{B}}}{100\ {\rm GeV}} \right),
  \label{eq:Ybino_focus}
\end{eqnarray}
where the yield variable is defined as $Y_x \equiv n_x/s$ with $n_x$
being the number density of particle $x$ and $s$ the entropy density
of the universe.

\begin{figure}[t]
 \centerline{\epsfxsize=0.55\textwidth\epsfbox{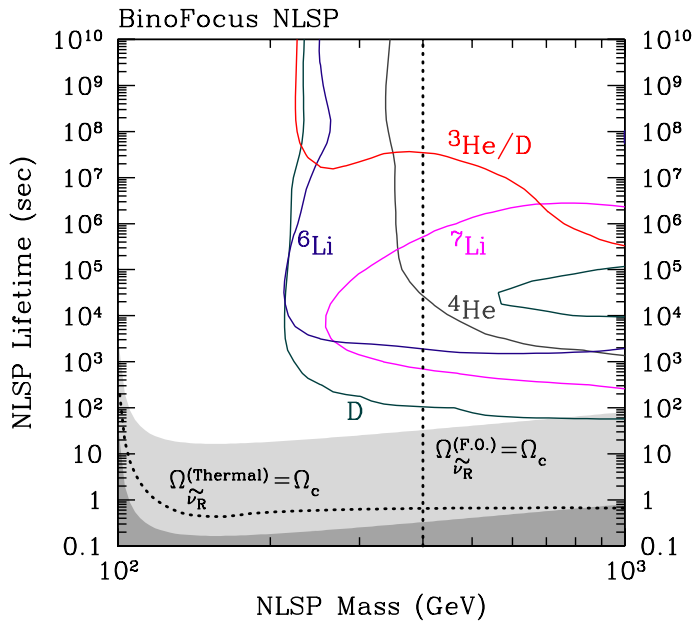}}
 \caption{\small BBN constraints on the Bino-NLSP case are shown on
   $m_{\tilde{B}}$ vs.\ $\tau_{\tilde{B}}$ plane.  The un-shaded,
   lightly shaded, and darkly shaded regions are for $a_\nu<1$,
   $1<a_\nu<10$, and $a_\nu>10$, respectively.  In addition, the
   contours of $\Omega^{\rm (F.O.)}_{\tilde{\nu}_R}=\Omega_c$ and
   $\Omega^{\rm (Thermal)}_{\tilde{\nu}_R}=\Omega_c$ are also shown
   (dotted lines).  For the calculation of $\Omega^{\rm
     (Thermal)}_{\tilde{\nu}_R}$, we take $m_{\tilde{\nu}_R}=100\ {\rm
     GeV}$, $m_{\tilde{\nu}_L}=1.2 m_{\tilde{B}}$,
   $\mu_H=2m_{\tilde{B}}$, $m_h=115~{\rm GeV}$, and $\tan \beta=30$.}
 \label{fig:const_Binofocus}
\end{figure}

\begin{figure}[t]
  \centerline{\epsfxsize=0.55\textwidth\epsfbox{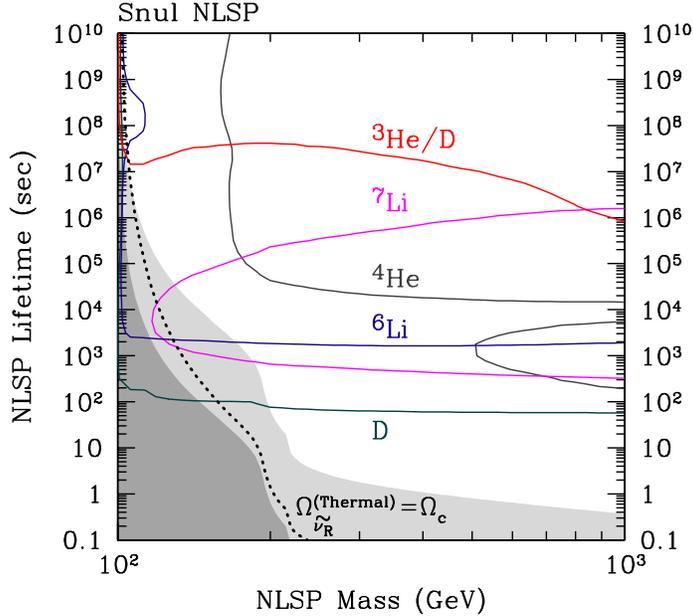}}
  \caption{\small BBN constraints on the $\tilde{\nu}_L$-NLSP case are
    shown on $m_{\tilde{\nu}_L}$ vs.\ $\tau_{\tilde{\nu}_L}$ plane.
    The un-shaded, lightly shaded, and darkly shaded regions are for
    $a_\nu<1$, $1<a_\nu<10$, and $a_\nu>10$, respectively.  In
    addition, the contour of $\Omega^{\rm
      (Thermal)}_{\tilde{\nu}_R}=\Omega_c$ is shown in dotted line.
    Here, we take $m_{\tilde{\nu}_R}=100\ {\rm GeV}$,
    $m_{\tilde{B}}=1.2 m_{\tilde{\nu}_L}$, $\mu_H=2m_{\tilde{B}}$,
    $m_h=115~{\rm GeV}$, and $\tan \beta=30$.}
  \label{fig:const_Snul}
\end{figure}

\begin{figure}[t]
 \centerline{\epsfxsize=0.55\textwidth\epsfbox{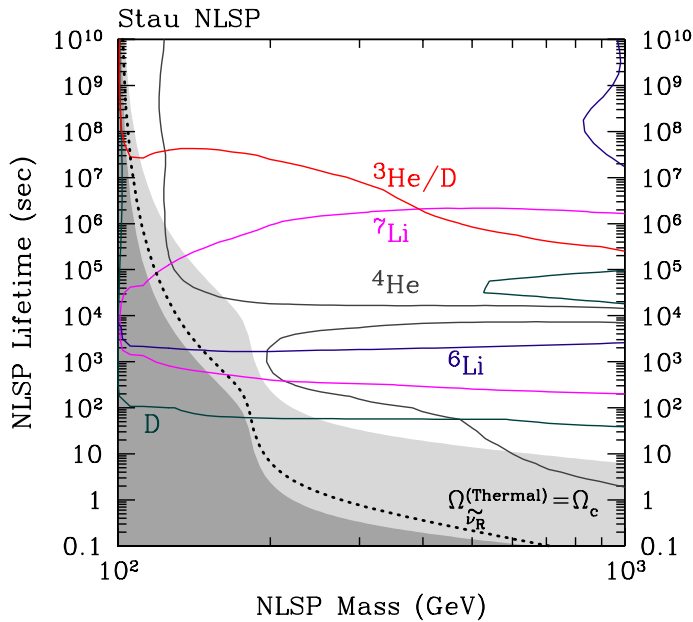}}
 \caption{\small BBN constraints on the $\tilde{\tau}$-NLSP case are
   shown on $m_{\tilde{\tau}}$ vs.\ $\tau_{\tilde{\tau}}$ plane.  The
   un-shaded, lightly shaded, and heavily shaded regions are for
   $a_\nu<1$, $1<a_\nu<10$, and $a_\nu>10$, respectively.  In
   addition, the contour of $\Omega^{\rm
     (Thermal)}_{\tilde{\nu}_R}=\Omega_c$ is shown in dotted line.
   Here, we take $m_{\tilde{\nu}_R}=100\ {\rm GeV}$,
   $m_{\tilde{B}}=1.2 m_{\tilde{\tau}}$, $m_{\tilde{\nu}_L}=1.2
   m_{\tilde{\tau}}$, $\mu_H=2m_{\tilde{B}}$, $m_h=115~{\rm GeV}$,
   $\tan \beta=30$, and $\sin\theta_{\tilde{\tau}}=0.3$.}
 \label{fig:const_Stau}
\end{figure}

Following the procedure given in \cite{KawKohMor}, we calculate the
light-element abundances taking account of the hadro-dissociation,
photo-dissociation, and $p\leftrightarrow n$ conversion processes.
The energy distribution of the final-state particles are calculated by
using the HELAS package \cite{Murayama:1992gi}, and the hadronization
processes of colored particles are studied by using the PYTHIA package
\cite{Sjostrand:2000wi}.  In the Bino-NLSP case, high energy neutrino
emitted by the Bino decay may scatter off background neutrino and
generate energetic $e^\pm$, which becomes the source of energetic
photon \cite{NeutrinoInjection}.  In our analysis, we have taken into
account the effects of the photo-dissociation process induced by
photon from the neutrino injection.  (However, we found that the
neutrino-induced processes are less important compared to other
processes.)  Once theoretical values of the primordial light-element
abundances are obtained as functions of the mass and the lifetime of
the NLSP, we compare them with the observed values of the primordial
abundances.  In deriving the constraints on the model, we adopt the
following observational constraints:
\begin{itemize}
\item D to H ratio \cite{O'Meara:2006mj,Amsler:2008zzb}:
  \begin{eqnarray}
    {\rm (n_{\rm D}/n_{\rm H})}_{\rm p} = 
    (2.82 \pm 0.26) \times 10^{-5}.
    \label{D/H}
  \end{eqnarray}
\item $^{4}$He mass fraction \cite{Izotov:2007ed,Fukugita:2006xy}:
  \begin{eqnarray}
    Y_{\rm p} = 0.2516 \pm 0.0040.
    \label{Yp}
  \end{eqnarray}
\item $^{3}$He to D ratio~\cite{GG03,KawKohMor}:
  \begin{eqnarray}
    (n_{\rm ^3He}/n_{\rm D})_{\rm p} < 0.83+0.27.
    \label{3He/D}
  \end{eqnarray}
\item $^{6}$Li to $^{7}$Li ratio
  \cite{Asplund:2005yt,Hisano:2009rc}:\footnote
  {Asplund {\it et al.}\ reported $n_{\rm ^6Li}/n_{\rm ^{7}Li} = 0.046
    \pm 0.022$. However, their positive detection has not been fully
    confirmed yet as pointed out in \cite{Perez:2009ax}.  Therefore,
    we consider the observed value as an upper bound.}
  \begin{eqnarray}
    ( n_{\rm ^6Li}/n_{\rm ^{7}Li} )_{\rm p} < 0.046 + 0.022 +
    0.106.
    \label{6Li/7Li}
  \end{eqnarray}
\item $^{7}$Li to H ratio \cite{Bonifacio:2006au,Hisano:2009rc}:
  \begin{eqnarray}
    \log_{10}(n_{^{7}{\rm Li}}/n_{\rm H})_{\rm p} 
    = -9.90 \pm 0.09 + 0.35.
    \label{7Li/H}
  \end{eqnarray}
\end{itemize}
(Here and hereafter the subscript ``p'' denotes the primordial value
inferred by observation.)  As shown in \eqref{6Li/7Li} and
\eqref{7Li/H}, we add positive systematic errors of $+0.106$ and
$+0.35$ to the observational face-values of $( n_{\rm ^6Li}/n_{\rm
  ^{7}Li} )_{\rm p}$ and $\log_{10}(n_{^{7}{\rm Li}}/n_{\rm H})_{\rm
  p}$, respectively.  We expect that these systematic errors result
from possible depletion in stars through rotational mixing
\cite{Pinsonneault:2001ub} or diffusion \cite{Korn:2006tv}.  Since
both $^7$Li and $^6$Li are destroyed by depletion process, their
systematic errors are correlated (for more details, see
\cite{Hisano:2009rc}).  We note here that the standard BBN is excluded
at more than 4-$\sigma$ level if we do not adopt the systematic error
on $^{7}$Li abundance \cite{Cyburt:2008kw} (so-called ${\rm ^7Li}$
problem).  Thus, to derive a conservative constraint, we add these
systematic errors.  At the end of this letter, we will comment on
implications of the $\tilde{\nu}_R$-DM scenario on the ${\rm ^7Li}$
problem.\footnote
{As we will discuss later in considering the $^{7}$Li problem, one may
  adopt a slightly higher value of D to H ratio, ${\rm (n_{\rm
      D}/n_{\rm H})}_{\rm p} = (3.98^{+0.59}_{-0.67}) \times 10^{-5}$
  \cite{O'Meara:2006mj}, and/or that of ${\rm ^7Li}$ to H ratio,
  $\log_{10}(n_{^{7}{\rm Li}}/n_{\rm H})_{\rm p} =-9.63 \pm 0.06$
  \cite{Melendez:2004ni}.  We have checked that, even with these
  observational constraints, the constraints given in Figs.\
  \ref{fig:const_Binofocus} $-$ \ref{fig:const_Stau} are almost
  unchanged (as far as the systematic error in the ${\rm ^7Li}$
  abundance is taken into account).}

In Fig.\ \ref{fig:const_Binofocus}, we show the constraint from BBN
for the Bino-NLSP case on $m_{\tilde{B}}$ vs.\ $\tau_{\tilde{B}}$
plane (with $\tau_{\tilde{B}}$ being the lifetime of Bino).  The
lifetime is related to the fundamental parameters via Eq.\
\eqref{Gamma_bino}; in particular, $\tau_{\tilde{B}}$ is proportional
to $a_\nu^{-2}$.  Taking $m_{\tilde{\nu}_L}=1.2 m_{\tilde{B}}$, we
calculate $a_\nu$-parameter.  In the figure, un-shaded, lightly
shaded, and darkly shaded regions indicate the region with $a_\nu<1$,
$1<a_\nu<10$, and $a_\nu>10$, respectively.  One can see that the
region with $m_{\tilde{B}} \lesssim 200~{\rm GeV}$ is always allowed.
This is because, in such a region, the dominant hadronic decay
processes are four-body ones ($\tilde{B}\rightarrow
\tilde{\nu}_R\bar{\nu} q\bar{q}$ and $\tilde{\nu}_R l q\bar{q}'$), for
which the branching ratio is significantly suppressed by the
phase-space factor.  On the other hand, when the decay processes
$\tilde{B}\rightarrow \tilde{\nu}_R\bar{\nu} Z$ and $\tilde{\nu}_R l
W$ are kinematically allowed, those three-body decay processes have
sizable branching ratio, resulting in an enhanced production of
hadrons.  We can see that the lifetime of Bino is constrained to be
smaller than $\tau_{\tilde{B}}\lesssim 10^2~{\rm sec}$ in such a
parameter region in order not to overproduce deuterium via the
hadro-dissociation of $^4$He.

In the figure, we also plot contours of constant density parameters.
Once the primordial abundance of the NLSP is fixed,
$\Omega^{\mbox{\tiny (F.O.)}}_{\tilde{\nu}_R}$ is calculated by using
Eq.\ \eqref{Omega(FO)}.  We show the contour of $\Omega^{\rm
  (F.O.)}_{\tilde{\nu}_R}=\Omega_c=0.228$ \cite{Hinshaw:2008kr}; the
right-hand side of the line is excluded by the overclosure constraint
if we adopt the abundance given in Eq.\ \eqref{eq:Ybino_focus}.  In
studying the $\tilde{\nu}_R$-DM scenario, we should also consider
$\tilde{\nu}_R$ from the MSSM particles in the thermal bath.
Following \cite{SnurLSP}, we calculate the sneutrino abundance by
solving the Boltzmann equation taking account of all the relevant
sneutrino production processes.  The contour of $\Omega^{\rm
  (Thermal)}_{\tilde{\nu}_R}=\Omega_c$ is shown in Fig.\
\ref{fig:const_Binofocus}; the $a_\nu$-parameter is determined by
using Eq.\ \eqref{Gamma_bino}, while the MSSM parameters are taken to
be $m_{\tilde{\nu}_L}=1.2 m_{\tilde{B}}$, and $\mu_H=2m_{\tilde{B}}$
(with $\mu_H$ being the SUSY invariant Higgs mass).  $\tilde{\nu}_R$
is overproduced below the line of $\Omega^{\rm
  (Thermal)}_{\tilde{\nu}_R}=\Omega_c$ with the present choice of
parameters.  One can see that the line is well below the constrained
region by BBN.  In the present choice of parameters, a relatively
large value of $a_\nu$ is needed unless the masses of $\tilde{B}$ and
$\tilde{\nu}_R$ are degenerate in order to realize $\Omega^{\rm
  (Thermal)}_{\tilde{\nu}_R}=\Omega_c$.  However, notice that the
relic abundance of $\tilde{\nu}_R$ depends on various parameters.  In
particular, $\Omega^{\rm (Thermal)}_{\tilde{\nu}_R}$ becomes larger
when the mass difference between $\tilde{\nu}_R$ and $\tilde{\nu}_L$
becomes smaller because the left-right mixing is enhanced.  In
addition, $\Omega^{\rm (Thermal)}_{\tilde{\nu}_R}$ is also enhanced if
we use a larger value of the neutrino Yukawa coupling constant; it may
happen when we adopt the degenerate neutrino masses.  Thus, with other
choices of parameters, the required value of $a_\nu$ to realize
$\Omega^{\rm (Thermal)}_{\tilde{\nu}_R}=\Omega_c$ changes.

Another candidate for the NLSP is the left-handed sneutrino.  Such a
scenario is attractive in the $\tilde{\nu}_R$-DM scenario because the
$\tilde{\nu}_R$ abundance is enhanced if the masses of $\tilde{\nu}_R$
and $\tilde{\nu}_L$ becomes closer.  When $\tilde{\nu}_L$ is the NLSP,
its dominant decay process is $\tilde{\nu}_L\rightarrow\tilde{\nu}_R
Z^{(*)}$ and $\tilde{\nu}_L\rightarrow\tilde{\nu}_R h^{(*)}$.  Thus,
colored and/or charged particles are effectively produced via the
dominant decay modes.  Again, we calculate the light-element
abundances taking account of the hadro-dissociation,
photo-dissociation, and $p\leftrightarrow n$ conversion processes, and
compare the resultant light-element abundances with observational
constraints given in \eqref{D/H} $-$ \eqref{7Li/H}.  The relic
abundance of left-handed sneutrino is approximated as
\cite{Fujii:2003nr}:
\begin{eqnarray}
  Y_{\tilde{\nu}_L} \simeq 2 \times 10^{-14} \times
    \left( \frac{m_{\tilde{\nu}}}{100\ {\rm GeV}} \right).
    \label{eq:Ysnul}
\end{eqnarray}

The BBN constraints are shown in Fig.\ \ref{fig:const_Snul}.  We can
see that the parameter space is constrained as $\tau_{\tilde{\nu}_L}
\lesssim 10^2~{\rm sec}$ (with $\tau_{\tilde{\nu}_L}$ being the
lifetime of $\tilde{\nu}_L$) by the deuterium overproduction
irrespective of $m_{\tilde{\nu}_L}$.  This is due to the fact that, if
$\tilde{\nu}_L$ is the NLSP, production of hadrons occurs in the
dominant decay processes.  This is a large contrast to the Bino-NLSP
case.

If $\tilde{\nu}_L$ is the NLSP, its primordial abundance is so small
that $\Omega_{\tilde{\nu}_R}^{\mbox{\tiny (F.O.)}}<\Omega_c$ as far as
$m_{\tilde{\nu}_L}\lesssim 10\ {\rm TeV}$ (for $m_{\tilde{\nu}_R}=100\
{\rm GeV}$).  On the contrary, $\Omega^{\rm
  (Thermal)}_{\tilde{\nu}_R}$ can be as large as $\Omega_c$; in the
figure, we plot the contour of $\Omega^{\rm
  (Thermal)}_{\tilde{\nu}_R}=\Omega_c$.  Here, the $a_\nu$-parameter
is determined for given values of $m_{\tilde{\nu}_L}$ and
$\tau_{\tilde{\nu}_L}$, while the MSSM parameters are taken to be
$m_{\tilde{B}} = 1.2m_{\tilde{\nu}_L}$, and $\mu_H=2m_{\tilde{B}}$.
One can see that, when $m_{\tilde{\nu}_L}\gtrsim 160\ {\rm GeV}$,
$\Omega^{\rm (Thermal)}_{\tilde{\nu}_R}=\Omega_c$ can be realized with
$a_\nu\lesssim 10$ (which is marginally consistent with the naive
order-of-estimate of the $a_\nu$-parameter in gravity-mediated SUSY
breaking scenario).  Notice that, even with $a_\nu\sim 1$ (or
smaller), $\Omega^{\rm (Thermal)}_{\tilde{\nu}_R}$ can be large enough
if a larger value of $y_\nu$ is adopted or if $\tilde{\nu}_R$ is
produced by the decay of some exotic particles.

Next, we consider the $\tilde{\tau}$-NLSP case.  In this case, because
the NLSP is charged, it may form a bound state with ${\rm ^4He}$
during the BBN epoch and change the reaction rate
\cite{Pospelov:2006sc}.  (Such an effect is called
$\tilde{\tau}$-catalyzed effect.)  Consequently, ${\rm ^6Li}$
abundance may be significantly enhanced if the lifetime of
$\tilde{\tau}$ is longer than $\sim 10^3\ {\rm sec}$.  Here, the
light-element abundances are calculated by including the
$\tilde{\tau}$-catalyzed effect.  Assuming that $\tilde{\tau}$ is
almost right-handed, we approximate the primordial abundance as
\cite{Fujii:2003nr}:
\begin{eqnarray}
  Y_{\tilde{\tau}} \simeq 7 \times 10^{-14} \times
  \left( \frac{m_{\tilde{\tau}}}{100\ {\rm GeV}} \right),
  \label{eq:Ystau}
\end{eqnarray}
and calculate the light-element abundance.  The numerical result is
shown in Fig.\ \ref{fig:const_Stau}.  As in the $\tilde{\nu}_L$-NLSP
case, the parameter space $\tau_{\tilde{\tau}} \gtrsim 10^2~{\rm sec}$
(with $\tau_{\tilde{\tau}}$ being the lifetime of $\tilde{\tau}$) is
excluded.  In addition, the ${\rm ^4He}$ is overproduced due to
$p\leftrightarrow n$ conversion process when $m_{\tilde{\tau}} \gtrsim
500\ {\rm GeV}$ and $\tau_{\tilde{\tau}}\sim 10\ {\rm sec}$.  The
${\rm ^4He}$ constraint becomes more stringent than the
$\tilde{\nu}_L$-NLSP case because the yield variable used in the
$\tilde{\tau}$-NLSP case is larger.  We also show the line which
satisfies $\Omega^{\rm (Thermal)}_{\tilde{\nu}_R}=\Omega_c$, taking
$m_{\tilde{B}}=1.2 m_{\tilde{\tau}}$, $m_{\tilde{\nu}_L}=1.2
m_{\tilde{\tau}}$, $\mu_H=2m_{\tilde{B}}$, and
$\sin\theta_{\tilde{\tau}}=0.3$.  As one can see, the lifetime becomes
longer for a given value of $a_\nu$ compared to the case of
$\tilde{\nu}_L$-NLSP; this is because we have taken a small value of
$\theta_{\tilde{\tau}}$.  Even in this case, we can see that
$\Omega^{\rm (Thermal)}_{\tilde{\nu}_R}=\Omega_c$ can be realized with
$a_\nu\lesssim 10$ in the parameter region consistent with all the BBN
constraints.

Finally we comment on the implication of the $\tilde{\nu}_R$-LSP
scenario on the so-called $^{7}$Li problem.  As we have mentioned, the
standard BBN is excluded at more than 4-$\sigma$ level if we take the
face value of the observational constraints on the $^{7}$Li abundance;
the theoretical prediction of the $^{7}$Li abundance becomes
significantly larger than the observed value.  Even though the
$^{7}$Li problem does not exist if a significant depletion of $^{7}$Li
occurs in stars, the degree of the depletion has not yet been
accurately understood.  If one adopts models with small depletion, the
astrophysical or particle-physics solution to the $^{7}$Li problem is
required.  It is notable that the $^{7}$Li abundance can be reduced if
a long-lived particle decays into hadrons during the BBN epoch
\cite{Jedamzik:2004er, Cumberbatch:2007me}.  Thus, in the present
case, the decay of the NLSP during the BBN may be a solution to the
$^{7}$Li problem.  In the following, we will see that the $^{7}$Li
problem may be solved if $\tilde{B}$ is the NLSP.  (For the cases of
$\tilde{\nu}_L$- and $\tilde{\tau}$-NLSP, the $^{7}$Li problem is
hardly solved because the parameter region with the lifetime longer
than $\sim 10^2\ {\rm sec}$ is (almost) excluded, as shown in Figs.\
\ref{fig:const_Snul} and \ref{fig:const_Stau}.)

To study the $^{7}$Li problem in the present framework, we neglect the
systematic error (i.e., $+0.35$ dex) in the observational constraint
on $^{7}$Li abundance.  In addition, because the allowed parameter
region is sensitive to the observational constraint on $^{7}$Li, we
consider two different observational constraints on $^{7}$Li abundance:
\begin{eqnarray}
  {\rm Low}~^{7}{\rm Li}:~\log_{10}(n_{^{7}{\rm Li}}/n_{\rm H})_{\rm p} 
  &=& -9.90 \pm 0.09~ \cite{Bonifacio:2006au},
  \label{7Li/H_2} \\
  {\rm High}~^{7}{\rm Li}:~\log_{10}(n_{^{7}{\rm Li}}/n_{\rm H})_{\rm p} 
  &=& -9.63 \pm 0.06~\cite{Melendez:2004ni}.
  \label{7Li/H_high}
\end{eqnarray}
Notice that the low value corresponds to the one given in
\eqref{7Li/H}, while the high value is from measurement using
different method to estimate temperature of the atmosphere in dwarf
halo stars.  In addition, because the systematic error in the $^{6}$Li
to $^{7}$Li ratio is correlated to that of $^{7}$Li, we also remove
the systematic error from \eqref{6Li/7Li}:
\begin{eqnarray}
  ( n_{\rm ^6Li}/n_{\rm ^{7}Li} )_{\rm p} < 0.046 + 0.022.
  \label{6Li/7Li(no-syst)}
\end{eqnarray}

\begin{figure}
  \centerline{\epsfxsize=0.5\textwidth\epsfbox{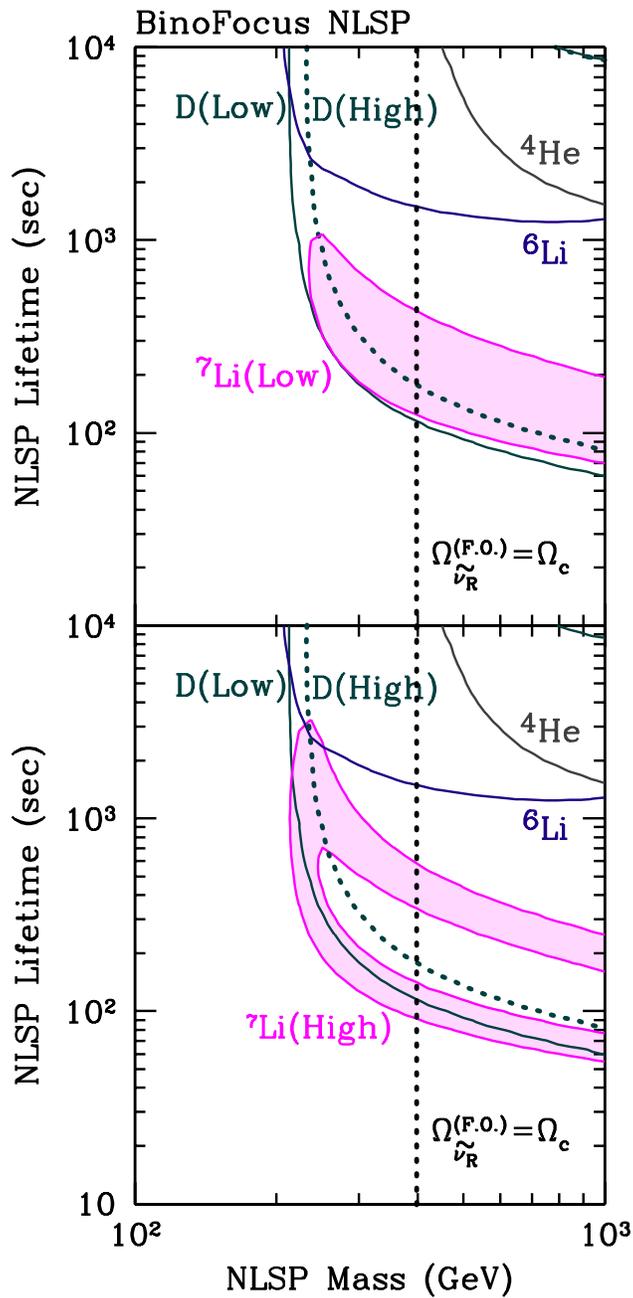}}
  \caption{Same as Fig.\ \ref{fig:const_Binofocus}, but with different
    set of observational constraints. Regions where ${\rm ^7Li}$
    abundance becomes consistent with the observation are shaded.}
  \label{fig:7LiProblem}
\end{figure}

BBN constraints on the Bino-NLSP case are shown in Fig.\
\ref{fig:7LiProblem}, using the constraints \eqref{7Li/H_2} and
\eqref{6Li/7Li(no-syst)} (upper panel) or \eqref{7Li/H_high} and
\eqref{6Li/7Li(no-syst)} (lower panel).  Here, constraints on ${\rm
  (n_{\rm D}/n_{\rm H})}_{\rm p}$, $Y_{\rm p}$, and $(n_{\rm
  ^3He}/n_{\rm D})_{\rm p}$ are unchanged from the previous cases; the
D to H ratio given in \eqref{D/H} is called ``Low D'' because of the
reason below.  As one can see, if we adopt the high value of the
$^{7}$Li to H ratio, all the light-element abundances can be
consistent with the observational constraints if $10^2\ {\rm
  sec}\lesssim\tau_{\tilde{B}}\lesssim 10^3\ {\rm sec}$.  On the
contrary, with the low value of ${\rm ^7Li}$ abundance, the constraint
on the D to H ratio \eqref{D/H} makes it difficult to solve the
$^{7}$Li problem.  However, this conclusion changes if we adopt a
slight systematic error in the ${\rm ^7Li}$ abundance, or if a
different observational constraint on the D to H ratio is adopted.
Indeed, in some literature, a higher value of the D to H ratio (which
is the highest value among the data points for six most precise
observations \cite{O'Meara:2006mj}) is adopted because D is the most
fragile light element and the observed values might reflect the
abundance after suffering from some destruction processes:
\begin{eqnarray}
    {\rm High~D:~(n_{\rm D}/n_{\rm H})}_{\rm p} &=& 
    (3.98^{+0.59}_{-0.67}) \times 10^{-5}.
    \label{D/H_High}
\end{eqnarray}
(We call this as ``High D.'')  In Fig.\ \ref{fig:7LiProblem}, we also
present the parameter region consistent with the constraint
\eqref{D/H_High} using the dotted line.  As one can see, with
\eqref{D/H_High}, the $^{7}$Li problem can be solved even with the low
value of the $^{7}$Li to H ratio.

Notice that, in the parameter region where all the light-element
abundances become consistent, $\Omega_{\tilde{\nu}_R}^{\mbox{\tiny
    (Thermal)}}$ becomes much smaller than $\Omega_c$ if the
constraint \eqref{7Li/H_2} or \eqref{7Li/H_high} is adopted.  However,
this fact does not imply that the $\tilde{\nu}_R$-LSP scenario cannot
solve the $^{7}$Li problem.  One possibility is to consider the
effects of the decay products of MSSM-LSP after freeze-out; indeed, as
shown in the figure, $\Omega_{\tilde{\nu}_R}^{\mbox{\tiny
    (F.O.)}}\simeq \Omega_c$ is realized when $m_{\tilde{B}}\sim 400\
{\rm GeV}$ and $\tau_{\tilde{B}}\sim 10^2\ {\rm sec}$ with solving the
$^{7}$Li problem.

\noindent {\it Acknowledgments:} This work was supported in part by
Research Fellowships of the Japan Society for the Promotion of Science
for Young Scientists (K.I.), and by the Grant-in-Aid for Scientific
Research from the Ministry of Education, Science, Sports, and Culture
of Japan, No. 14102004 (M.K.), No.\ 18071001 (K.K.) and No.\ 19540255
(T.M.), and also by World Premier International Research Center
Initiative, MEXT, Japan (M.K. and T.M.).


\begin{thebibliography}{99}

\bibitem{Hinshaw:2008kr}
  G.~Hinshaw {\it et al.}  [WMAP Collaboration],
  Astrophys.\ J.\ Suppl.\  {\bf 180}, 225 (2009).

\bibitem{SnurLSP}
  T.~Asaka, K.~Ishiwata and T.~Moroi,
  Phys.\ Rev.\  D {\bf 73}, 051301 (2006);
  Phys.\ Rev.\  D {\bf 75}, 065001 (2007).

\bibitem{Hall:2009bx}
  L.~J.~Hall, K.~Jedamzik, J.~March-Russell and S.~M.~West,
  arXiv:0911.1120 [hep-ph].

\bibitem{Moroi:1993mb}
  T.~Moroi, H.~Murayama and M.~Yamaguchi,
  Phys.\ Lett.\  B {\bf 303}, 289 (1993).

\bibitem{SuperWIMP}
  J.~L.~Feng, A.~Rajaraman and F.~Takayama,
  Phys.\ Rev.\ Lett.\  {\bf 91}, 011302 (2003);
  Phys.\ Rev.\  D {\bf 68}, 063504 (2003).

\bibitem{Ishiwata:2007bt}
  K.~Ishiwata, S.~Matsumoto and T.~Moroi,
  Phys.\ Rev.\  D {\bf 77}, 035004 (2008).

\bibitem{VariousSnuDMPapaers}
  S.~Gopalakrishna, A.~de Gouvea and W.~Porod,
  JCAP {\bf 0605}, 005 (2006);
  J.~March-Russell, C.~McCabe, M.~McCullough,
  arXiv:0911.4489 [hep-ph].

\bibitem{Ahn:2006zza}
  M.~H.~Ahn {\it et al.}  [K2K Collaboration],
  Phys.\ Rev.\  D {\bf 74}, 072003 (2006).

\bibitem{Feng:2004mt}
  J.~L.~Feng, S.~Su and F.~Takayama,
  Phys.\ Rev.\  D {\bf 70}, 075019 (2004).

\bibitem{KawKohMor}
  M.~Kawasaki, K.~Kohri and T.~Moroi,
  Phys.\ Lett.\  B {\bf 625}, 7 (2005);
  Phys.\ Rev.\  D {\bf 71}, 083502 (2005).

\bibitem{Murayama:1992gi}
  H.~Murayama, I.~Watanabe and K.~Hagiwara,
  ``HELAS: HELicity amplitude subroutines for Feynman diagram evaluations,''
  KEK-91-11.

\bibitem{Sjostrand:2000wi} 
  T.~Sjostrand {\it et al.},
  Comput.\ Phys.\ Commun.\  {\bf 135}, 238 (2001).

\bibitem{NeutrinoInjection}
  M.~Kawasaki and T.~Moroi,
  Phys.\ Lett.\  B {\bf 346}, 27 (1995);
  T.~Kanzaki, M.~Kawasaki, K.~Kohri and T.~Moroi,
  Phys.\ Rev.\  D {\bf 76}, 105017 (2007).

\bibitem{O'Meara:2006mj}
  J.~M.~O'Meara {\it et al.},
  Astrophys. J. {\bf 649}, L61 (2006).

\bibitem{Amsler:2008zzb}
  C.~Amsler {\it et al.}  [Particle Data Group],
  Phys.\ Lett.\  B {\bf 667}, 1 (2008).

\bibitem{Izotov:2007ed}
  Y.~I.~Izotov, T.~X.~Thuan and G.~Stasinska,
  arXiv:astro-ph/0702072.

\bibitem{Fukugita:2006xy}
  M.~Fukugita and M.~Kawasaki,
  Astrophys.\ J.\  {\bf 646}, 691 (2006).

\bibitem{GG03}
  J. Geiss and  G. Gloeckler, 
  Space Sience Reviews {\bf 106}, 3 (2003).

\bibitem{Asplund:2005yt}
     M.~Asplund {\it et al.},
     Astrophys.\ J.\  {\bf 644}, 229 (2006).

\bibitem{Hisano:2009rc}
  J.~Hisano {\it et al.},
  Phys.\ Rev.\  D {\bf 79}, 083522 (2009).

\bibitem{Perez:2009ax}
  A.~E.~G.~Perez {\it et al.},
  arXiv:0909.5163 [astro-ph.SR].

\bibitem{Bonifacio:2006au}
  P.~Bonifacio {\it et al.},
  arXiv:astro-ph/0610245.

\bibitem{Pinsonneault:2001ub}
  M.~H.~Pinsonneault, T.~P.~Walker, G.~Steigman and V.~K.~Narayanan,
  Astrophys.\ J.\  {\bf 527}, 180 (2002);
  M.~H.~Pinsonneault, G.~Steigman, T.~P.~Walker and V.~K.~Narayanans,
  Astrophys.\ J.\  {\bf 574}, 398 (2002).

\bibitem{Korn:2006tv}
  A.~J.~Korn {\it et al.},
  Nature {\bf 442}, 657 (2006).

\bibitem{Cyburt:2008kw}
  R.~H.~Cyburt, B.~D.~Fields and K.~A.~Olive,
  JCAP {\bf 0811}, 012 (2008);
R.~H.~Cyburt and B.~Davids,
  Phys.\ Rev.\  C {\bf 78}, 064614 (2008).

\bibitem{Melendez:2004ni}
  J.~Melendez and I.~Ramirez,
  Astrophys.\ J.\  {\bf 615}, L33 (2004).

\bibitem{Fujii:2003nr}
  M.~Fujii, M.~Ibe and T.~Yanagida,
  Phys.\ Lett.\  B {\bf 579}, 6 (2004).

\bibitem{Pospelov:2006sc}
  M.~Pospelov,
  Phys.\ Rev.\ Lett.\  {\bf 98}, 231301 (2007).

\bibitem{Jedamzik:2004er}
  K.~Jedamzik,
  Phys.\ Rev.\  D {\bf 70}, 063524 (2004).

\bibitem{Cumberbatch:2007me}
  D.~Cumberbatch {\it et al.},
  Phys.\ Rev.\  D {\bf 76}, 123005 (2007).

\end{thebibliography}
\end{document}